\documentstyle[12pt,epsf]{article}
\begin{document}

\begin{titlepage}
\begin{flushright}
\vspace{-2.0cm}{\normalsize UTHEP-304\\
UTCCP-P-1\\
May 1995\\}
\end{flushright}

\vspace*{3.0cm}
\begin{centering}
{\Large \bf QCD Phase Transition with Strange Quark\\
\vspace*{0.5cm}
in Wilson Formalism for Fermions}

\vspace{2.0cm}

{\large 
Y.\ Iwasaki\rlap,$^{\rm a,b}$ K.\ Kanaya\rlap,$^{\rm a,b}$ 
S.\ Kaya\rlap,$^{\rm a}$ S.\ Sakai\rlap,$^{\rm c}$ 
and T.\ Yoshi\'e$^{\rm a,b}$
}
\vspace{0.5cm}

{\it
$^{\rm a}$
Institute of Physics, University of Tsukuba,
Ibaraki 305, Japan\\
$^{\rm b}$
Center for Computational Physics, University of Tsukuba, \\
Ibaraki 305, Japan\\
$^{\rm c}$
Faculty of Education, Yamagata University,
Yamagata 990, Japan
}

\end{centering}

\vspace{2.0cm}\noindent
{
The nature of QCD phase transition is studied with massless 
up and down quarks and a light strange quark, using the Wilson 
formalism for quarks on a lattice with the temporal direction 
extension $N_t=4$.  We find that the phase transition is first 
order in the cases of both about 150 MeV and 400 MeV for the 
strange quark mass.  These results together with those for three 
degenerate quarks suggest that QCD phase transition in nature is 
first order.
}

\vfill \noindent

\end{titlepage}



One of major goals of numerical studies of lattice QCD 
is to determine the nature
of the phase transition from the high temperature
quark-gluon-plasma phase to the low temperature hadron 
phase, which is supposed to occur at the early stage of 
the Universe and possibly at heavy ion collisions.
It is, in particular, crucial to know
the order of the transition to understand the evolution 
of the Universe. 
Because the critical temperature is of the same order 
of magnitude as the strange quark mass, 
we have to take into account 
the effect of the strange quark 
as well as those of almost massless up and down quarks.
In this article we study 
the nature of QCD phase transition with these three quarks
in the Wilson formalism of fermions on the lattice.
The investigation using Wilson quarks is highly important,
because the Wilson formalism of fermions on the lattice
is the only known formalism which possesses a local action 
for any number of flavors with arbitrary quark masses.
Preliminary reports are given in \cite{preliminary}.

The results of numerical simulations with Wilson quarks 
\cite{ours} as well as with staggered quarks \cite{KSfl}
imply that the chiral phase transition is second order 
for two massless quarks,
while it is first order for three. 
This is also consistent with theoretical predictions 
based on universality \cite{Wilczek}.
In a numerical study we are able to vary the mass of the 
strange quark.
When the mass of the strange quark
is reduced from infinity to zero, the phase transition must 
change from second
order to first order at some quark mass $m_s^*$.
This is a tricritical point. The crucial point is whether
the physical strange quark mass is larger or smaller 
than $m_s^*$.


The action we use in this article is composed of 
the standard one-plaquette
gauge action with the gauge coupling 
constant $g$ ($\beta=6/g^2$) and the Wilson quark actions
for u, d and s quarks with the
hopping parameters $K_u$, $K_d$ and $K_s$, respectively.
We set $K_u=K_d$.

We perform simulations on lattices of the temporal extension 
$N_t=4$ with the spatial sizes $8^2 \times 10$ and $12^3$.
Although $N_t=4$ is not large enough to obtain
the result for the continuum limit,
this work is a first step toward the understanding of the 
nature of the QCD phase transition with Wilson quarks.
We generate gauge configurations by the hybrid R algorithm 
with a molecular dynamics step $\Delta \tau =0.01$ and a 
trajectory of unit time.
The inversion of the quark matrix 
is made by a minimal residual method or a conjugate gradient
method.
When the hadron spectrum is calculated,
the lattice is duplicated in the direction of lattice size
10 or 12.
We use an anti-periodic boundary  condition for quarks in the
$t$ direction and periodic boundary conditions otherwise.
The statistics is in general total $\tau=$ several hundreds,
and the plaquette and the Polyakov loop are measured every
simulation time unit and
hadron spectrum is calculated every $\delta\tau=5$ 
(or 10 sometimes).
When the value of $\beta$ is small, the fluctuation of
physical quantities are small
and therefore we think the statistics
is sufficient for the purpose of this article.

\begin{figure}
\centerline{ \epsfxsize=10.3cm \epsfbox{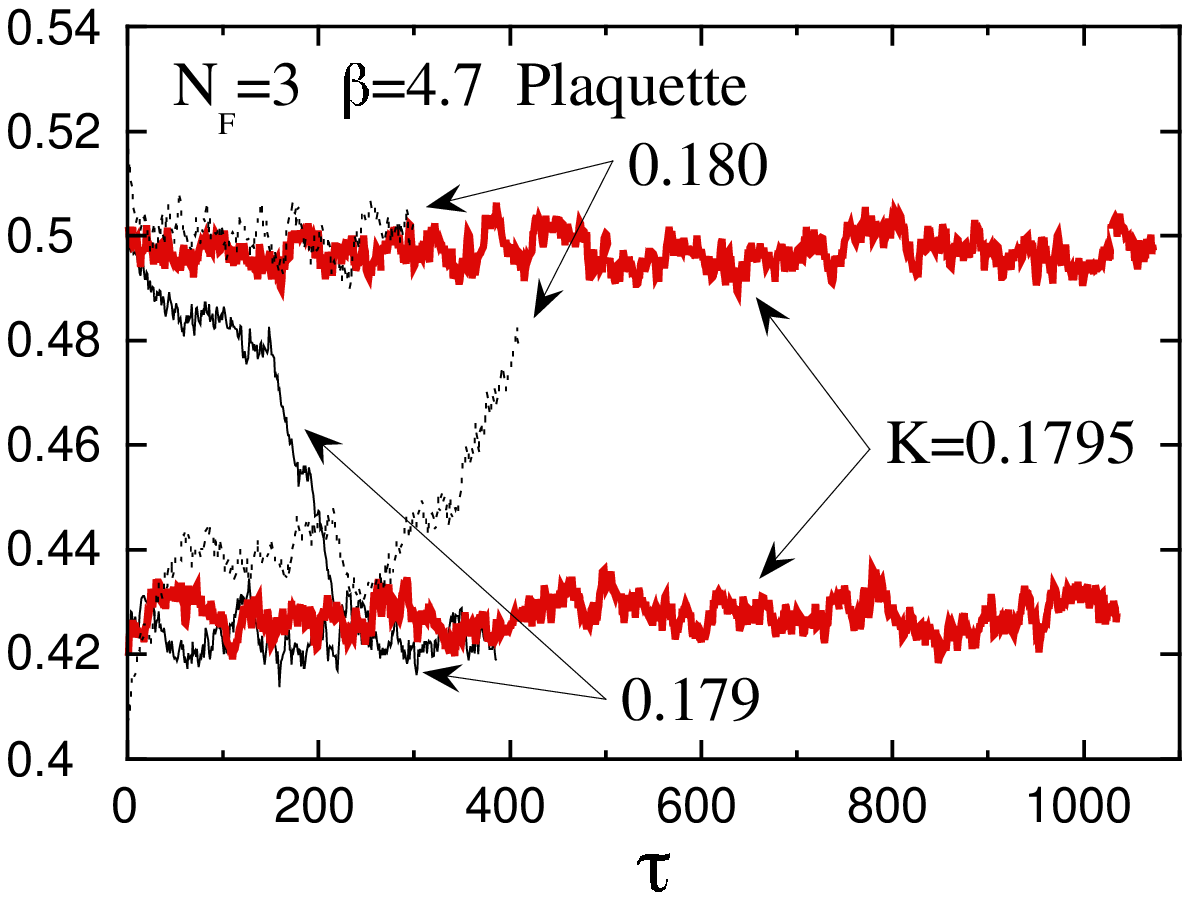} }
\centerline{\makebox[2mm]{} \epsfxsize=10.5cm \epsfbox{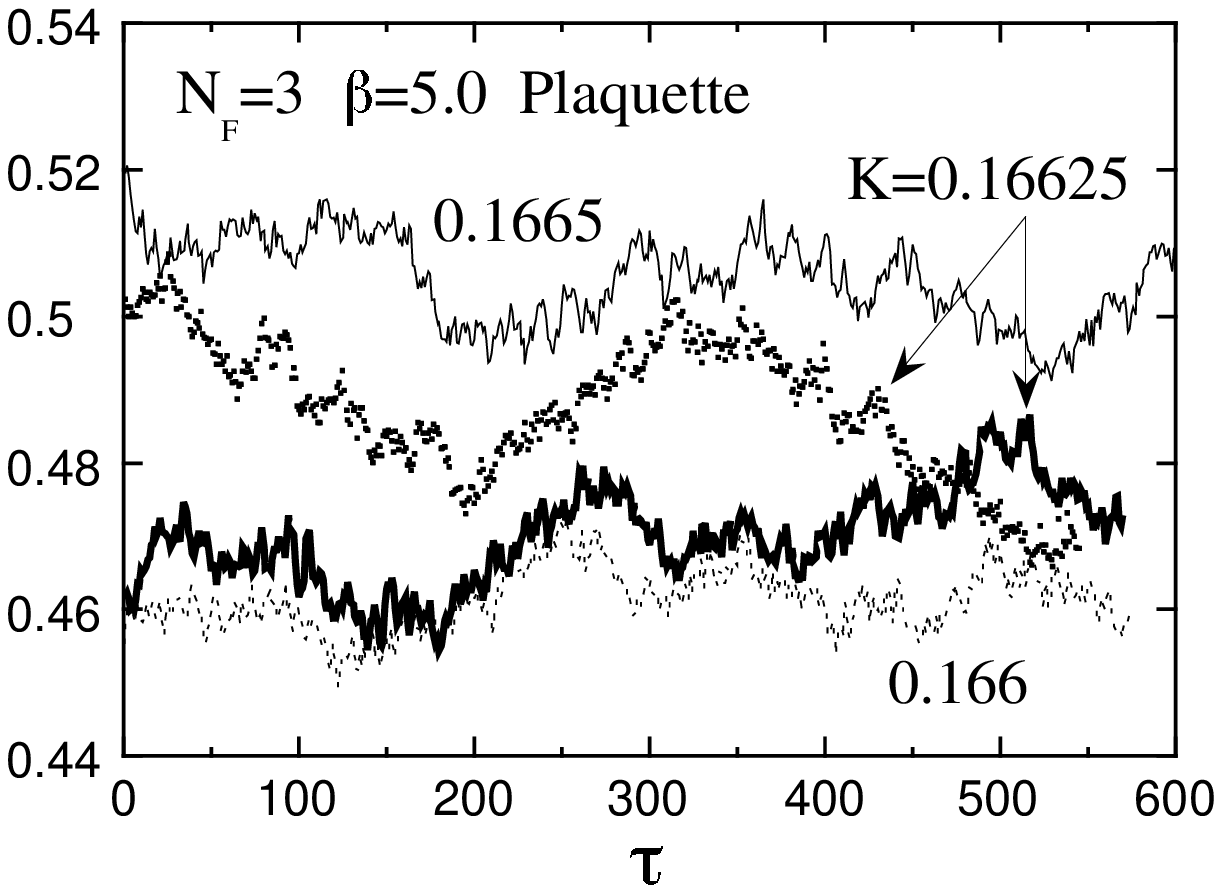}}
\caption{{
Time history of the plaquette for $N_F=3$ at (a) $\beta=4.7$ 
and (b) 5.0 on a $12^3\times4$ lattice.
}
\label{fig:H3W}}
\end{figure}
\begin{figure}
\centerline{ \epsfxsize=13cm \epsfbox{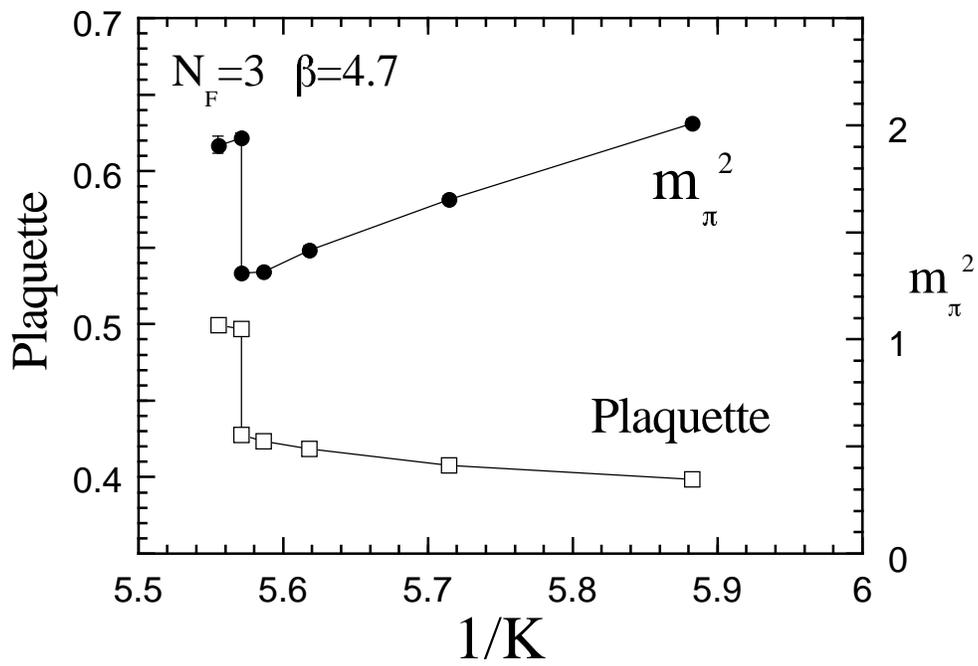} }
\caption{{
$(m_{\pi} a)^2$ and the plaquette for $N_F=3$ at $\beta=4.7$.
}
\label{fig:F3Pi}}
\end{figure}

Let us first discuss the results for the degenerate $N_F=3$ 
case:  $K_{u}=K_{d}=K_{s}\equiv K$.
In order to find the transition/crossover points 
we perform simulations at $\beta$=4.0, 4.5, 4.7, 5.0 and 5.5.
The simulation time history of the plaquette 
at $\beta=4.7$ on a $4 \times 12^3$ lattice is plotted in 
Fig.~\ref{fig:H3W}(a). 
The low and high temperature phases coexist over 1,000 
trajectories at $K=0.1795$ and in accord with this
we find two-state signals also in other observables such as 
the plaquette and $m_\pi^2$
(Fig.~\ref{fig:F3Pi}). 
From them we conclude that the transition at $K=0.1795(5)$ and
$\beta=4.7$ is first order.
The transition/crossover points thus identified
are summarized in Table 1.

\begin{figure}
\centerline{ \epsfxsize=12cm \epsfbox{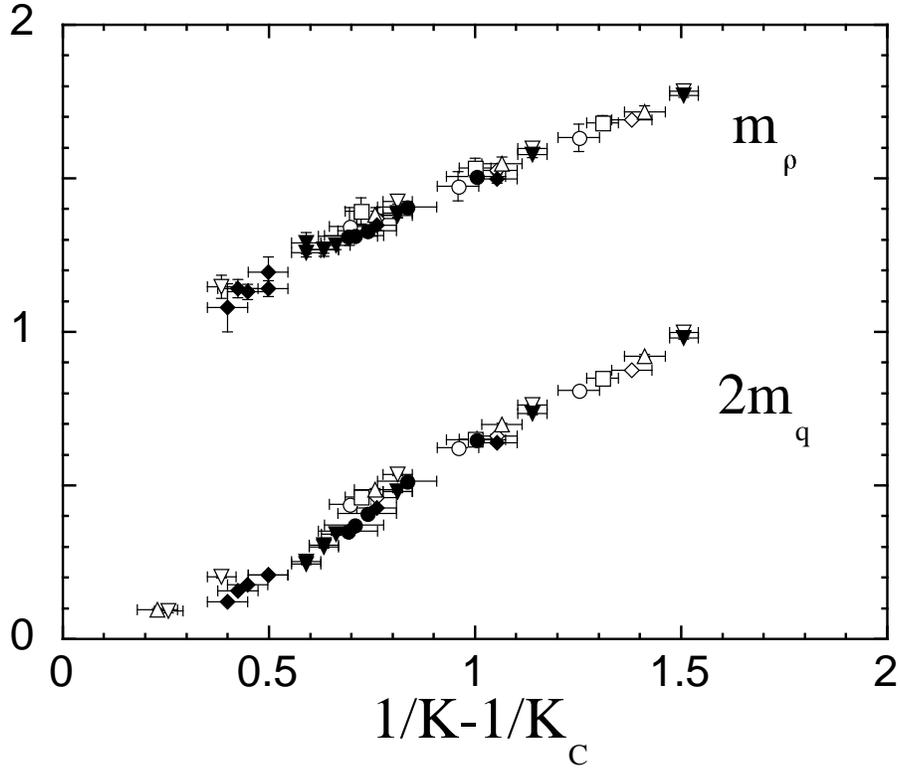} }
\caption{{
$m_{\rho}a$ and $2m_q a$ in the low temperature phase. 
Open symbols are for $N_F=2$, $\beta =  3.0$, 
3.5, 4.0, 4.3, and 4.5 on an $8^2\times10\times4$ lattice. 
Filled symbols are for $N_F=3$, $\beta=4.0$, 4.5 and 4.7 on 
$8^2\times10\times4$ and $12^3\times4$ lattices. 
$K_c$ for $N_F=2$ is used. 
Horizontal errors are from those of 
$K_c$ with taking into account the difference due to 
definitions, either the vanishing point
of $m_{\pi}^2$ or $m_q$.
}
\label{fig:F2F3Rho}}
\end{figure}

In a previous paper \cite{ours} we reported 
a clear two-state signal in the chiral limit 
$K_c=0.235$ at $\beta = 3.0$ for $N_F=3$.
An interesting question is to determine 
the critical value
of the quark mass $m_q^{crit}$ up to which
the first order phase transition persists.
We observed clear two state signals at 
$\beta$ = 4.0, 4.5 and 4.7, while 
for $\beta=5.0$ and $5.5$ no such signals have been seen 
(see Fig.~\ref{fig:H3W}(b) for the time history of the 
plaquette at $\beta=5.0$).
At the transition point (in the confining phase) of $\beta=4.7$
the value of $m_q a$ is 0.175(2) and
$(m_{\pi}/m_{\rho})= 0.873(6)$.
The quark mass is defined through the axial Ward identity 
\cite{Ward} for Wilson quarks.
The results of the hadron spectrum in the range of $\beta = 3.0$ 
-- 4.7 for $N_F=2$ and 3 (Fig.~\ref{fig:F2F3Rho}) indicate that 
the inverse lattice spacing $a^{-1}$
estimated from the $\rho$ meson mass is almost independent on 
$\beta$ in this range and $a^{-1} \sim 0.8$ GeV.
Therefore we obtain a bound on the critical quark mass 
$m_q^{crit} 
\mathop{\vtop{\ialign{#\crcr
$\hfil\displaystyle{>}\hfil$\crcr
\noalign{\kern0.5pt\nointerlineskip}
$\sim$\crcr\noalign{\kern0.5pt}}}}\limits 
140$ MeV,
or equivalently $(m_{\pi}/m_{\rho})^{crit} \ge 0.873(6)$.

We note 
that these values are much larger than those for 
the staggered fermion case for which
$m_q^{crit} \sim 10$ -- 40 MeV
($m_q^{crit} a=0.025$ -- 0.075 \cite{Gavai,Columbia}. 
We have used $a^{-1} \sim 0.5$ GeV at $\beta=5.2$ for 
$N_F$=2\cite{ukawa93}) 
and $(m_{\pi}/m_{\rho})^{crit} \simeq 0.42$ -- 0.58
(the result for $N_F=4$ at $\beta=5.2$\cite{Born89}, 
because the data for $N_F=3$ are not available.)


Now let us discuss a more realistic case of 
massless u and d quarks and a light s quark ($N_F=2+1$).
Our strategy to investigate this problem 
is similar to that
we utilized in our previous work \cite{ours}
for the investigation of the chiral
transition in the degenerate quark mass cases,
which we call ``on-$K_c$'' simulation method.
We set the value of the masses of u and d quarks $m_{ud}$  
to zero ($K_u=K_d= K_c$)
and fix the s quark mass $m_s$ to some value,
and make simulations reducing the value of $\beta$ starting 
from a deconfining phase.
When u and d quarks are massless, the number of iteration 
$N_{\rm inv}$
needed for the quark matrix inversion (for u and d quarks) 
is enormously large in the confining phase, while it is of 
order of several
hundreds in the deconfining phase. 
This is due to the fact that
there are zero modes around $K_c$ in the low temperature 
phase,
while none exists in the high temperature
phase \cite{strong}.
Therefore the $N_{\rm inv}$ provides an extremely good indicator
to discriminate the two phase when u and d quarks are on the 
$K_c$ line.
Combining this with measurements of physical observables, 
we identify the transition point
and determine the order of the transition.

The values which we take for $K_c$ are given in Table 2.
They are the vanishing point of extrapolated $m_\pi^2$ for 
$N_F=2$ and interpolated ones.
We have used those for $N_F=2$, because we have the data most 
in this case,
and the difference between that for $N_F=3$ and for $N_F=2$ is
of the same order of magnitude as the difference
due to the definition of $K_c$, either the
vanishing point of $m_q$ or $m_\pi^2$ \cite{ours}.

We study two cases of
$m_s \sim 150$ MeV and 400 MeV. 
From the value of $a^{-1} \sim 0.8$ GeV and 
an empirical rule $m_q a \simeq (2/3)(1/K-1/K_c)$ 
satisfied in the $\beta$ region we have studied
(Fig.~\ref{fig:F2F3Rho}),
we get the values for $K_s$ shown in Table 2.
It should be noted that the physical s quark mass 
determined from $m_\phi =1020$ MeV
turns out to be
$m_s \sim 150$ MeV in this $\beta$ range
for our definition of the quark mass.

\begin{figure}
\centerline{ \epsfxsize=10.5cm \epsfbox{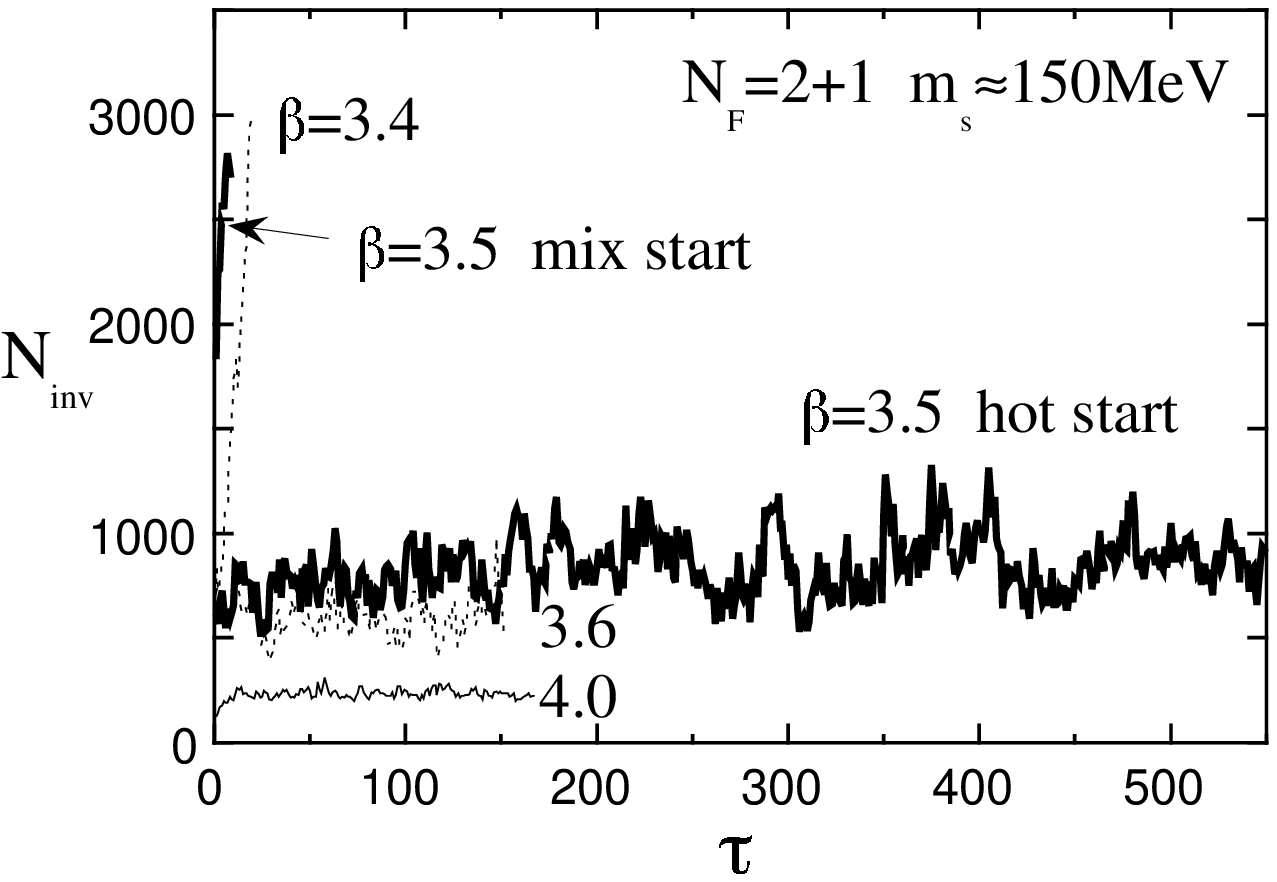} }
\centerline{\makebox[2mm]{} \epsfxsize=10.3cm \epsfbox{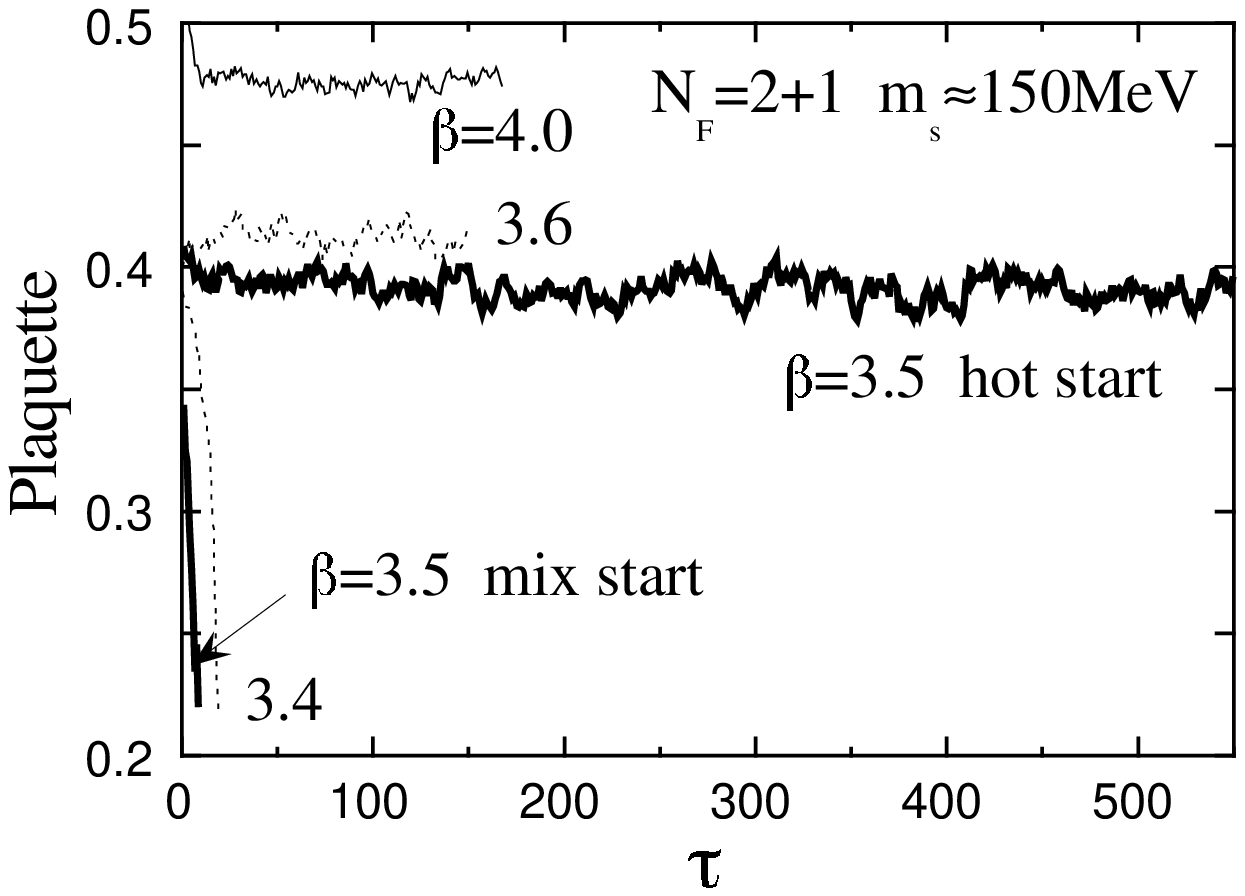}}
\caption{{
Time history of (a) $N_{\rm inv}$ and (b) the plaquette for 
$m_s \sim 150$ MeV on an $8^2\times10\times4$ lattice.
}
\label{fig:H21S150}}
\end{figure}

\begin{figure}
\centerline{ \epsfxsize=12cm \epsfbox{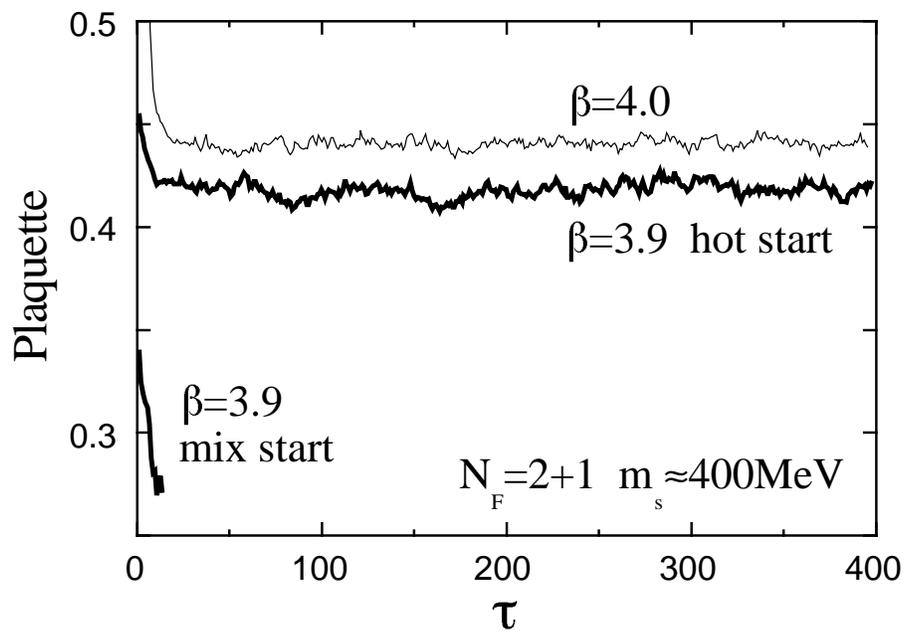} }
\caption{{
Time history of the plaquette for 
$m_s \sim 400$ MeV on a $12^3\times4$ lattice.
}
\label{fig:H21S400}}
\end{figure}

\begin{figure}
\centerline{ \epsfxsize=12cm \epsfbox{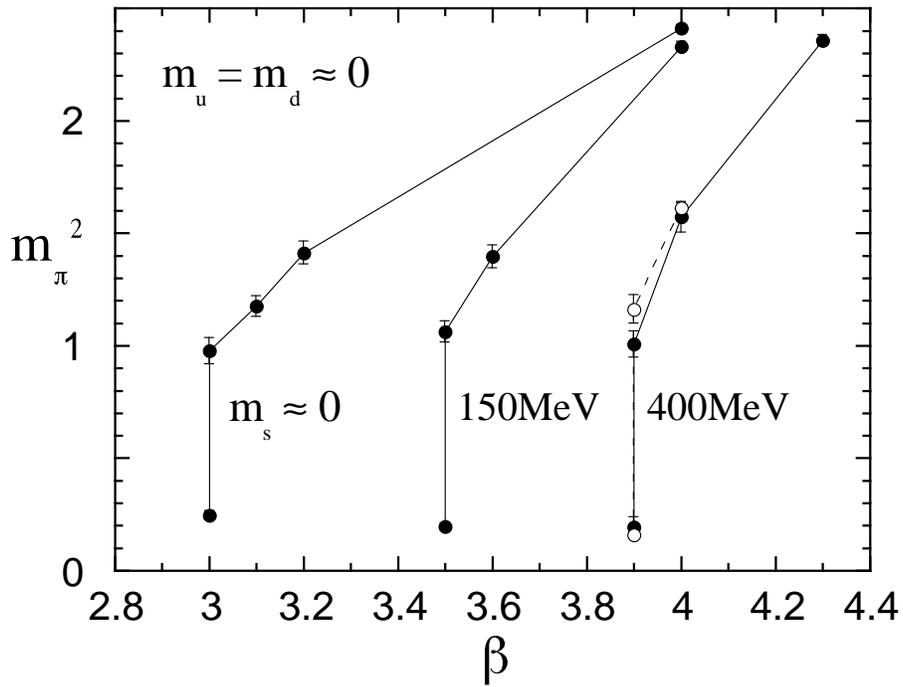} }
\caption{{
$(m_{\pi}a)^2$ versus $\beta$ for $m_s \simeq 0$, 150 and 400 
MeV with $m_{ud} \simeq 0$. 
Filled and open symbols are for $8^2 \times 10 \times 4$ 
and $12^3 \times 4$ lattices, respectively. 
}
\label{fig:F321Pi}}
\end{figure}

The simulation time history of $N_{\rm inv}$ on the 
$8^2 \times 10$ spatial lattice is plotted in 
Fig.~\ref{fig:H21S150}(a) for the case of 150 MeV.
When $\beta \ge 3.6$, $N_{\rm inv}$ is of order of several
hundreds, while when $\beta \le 3.4$, $N_{\rm inv}$
shows a rapid increase with $\tau$.
At $\beta = 3.5$ we see a clear two-state signal
depending on the initial condition:
For a hot start, $N_{\rm inv}$
is quite stable around $\sim 900$ and $m_\pi^2$ is large
($\sim 1.0$ in lattice units).
On the other hand, for a mix start, $N_{\rm inv}$
shows a rapid increase with $\tau$ and exceeds
2,500 in $\tau \sim 10$, and in accord with this, 
the plaquette and $m_\pi^2$
decreases with $\tau$ as shown in Fig.~\ref{fig:H21S150}(b) 
for the plaquette. 
For the case of 400 MeV a similar clear two-state signal 
is observed at $\beta=3.9$
both on the $8^2 \times 10$ and $12^3$ spatial lattices 
(Fig.~\ref{fig:H21S400}). 
Thus the method of ``on-$K_c$'' simulation 
is very powerful to identify the point of the transition,
although in the low temperature phase 
we are only able to obtain upper (or lower) bounds of physical 
observables. 
The values of $m_\pi^2$ versus $\beta$ are plotted in 
Fig.~\ref{fig:F321Pi} together with those in 
the case of degenerate $N_F=3$ on the $K_c$ line \cite{ours}.
At $\beta =3.5$ for the case of 150 MeV and at $\beta=3.9$ for 
the case of 400 MeV,
we have two values for $m_\pi^2$ depending
on the initial configuration.
The larger ones of order 1.0 are for hot starts, 
while the smaller ones are upper bounds for 
mix starts.

It should be noted that
the critical $\beta$ 
value of 3.9 for the case of $m_s \sim 400$ MeV is close to 
that of the chiral transition for $N_F=2$ at 
$\beta \sim  3.9$ -- 4.0 \cite{ours}. 
When $m_s$ is larger enough than 
the inverse lattice spacing,
the transition point should become almost identical with that 
of $N_F=2$
and the transition should change from first order to continuous.
Therefore the closeness of the two critical points
is quite natural, because the value of 400 MeV is not far from
that of the inverse lattice spacing of about 800 MeV.

\begin{figure}
\centerline{ \epsfxsize=12cm \epsfbox{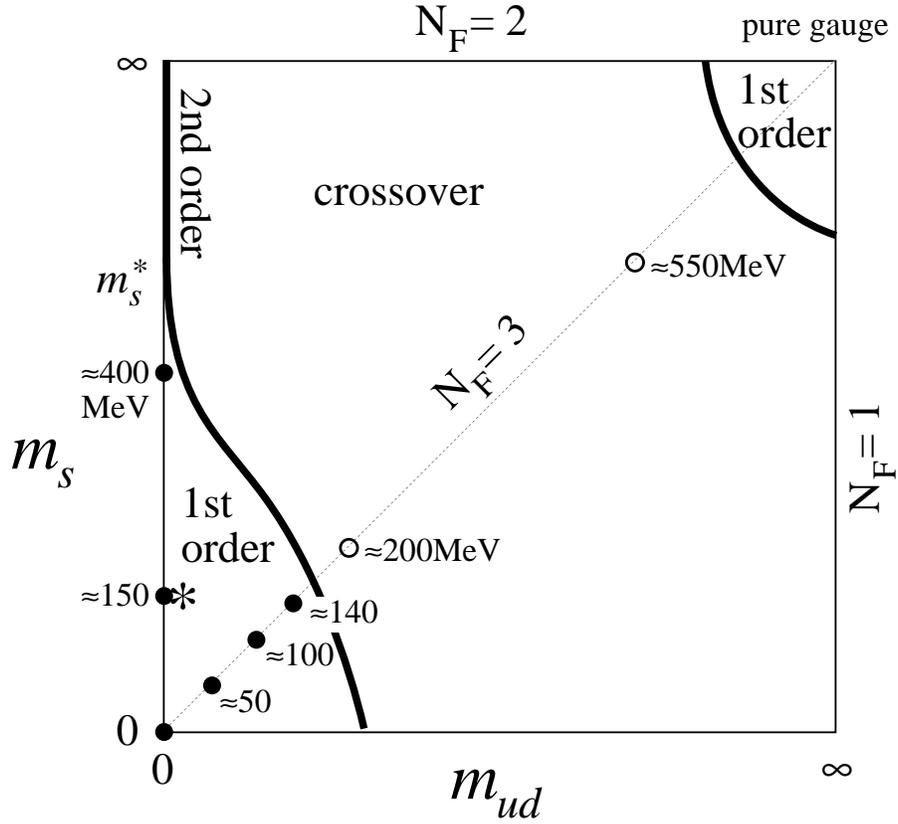} }
\caption{{
Order of the finite temperature QCD transition in the 
$(m_{ud},m_s)$ plane.
First order signals are observed at the points marked with 
filled circle, while no clear two state signals are found 
at the points with open circle. 
The values of quark mass in physical units are computed 
using $a^{-1} \sim 0.8$ GeV.
The real world corresponds to the point marked with star. 
The second order transition line is suggested 
to deviate from 
the vertical axis as $m_{ud} \propto (m_s^{*} - m_s)^{5/2}$
below $m_s^{*}$.
}
\label{fig:MsMud}}
\end{figure}

The above results imply that
$$m_s^* 
\mathop{\vtop{\ialign{#\crcr
$\hfil\displaystyle{>}\hfil$\crcr
\noalign{\kern0.5pt\nointerlineskip}
$\sim$\crcr\noalign{\kern0.5pt}}}}\limits
400\, {\rm MeV}.$$
Our results for the presence or absence of the transition in 
the cases of degenerate $N_F=3$ and $N_F=2+1$ are summarized 
in Fig.~\ref{fig:MsMud}.
Clearly the point which corresponds to the physical values of 
the u,d and s quarks 
exists in the range of the first order transition.
Thus our results suggest 
that the transition for the physical quark masses is of 
first order.

The result by the Columbia group \cite{Columbia} for 
staggered quarks shows that no transition occurs
at $m_u a = m_d a = 0.025$, $m_s a = 0.1$
($m_u=m_d \sim 12$ MeV, $m_s \sim 50$ MeV using 
$a^{-1} \sim 0.5$ GeV), which suggests that 
a first-order phase transition does not occur 
for the physical u,d and s quarks.
This result is in conflict with ours,
albeit with slightly heavier masses for u and d quarks.
One possibility for the discrepancy
is that both/either are far from the continuum limit.
Consistency between Wilson and
staggered quarks is an issue which should be
investigated in future.

The simulations are
performed with HITAC S820/80 at KEK
and with VPP500/30 at the University of Tsukuba.
We would like to thank members of KEK
for their hospitality and strong support.
This project is in part supported by the Grant-in-Aid
of Ministry of Education, Science and Culture
(No.06NP0401).


\clearpage

\begin{table}
\begin{center}
\begin{tabular}{cc}
\hline
$\beta$ & $K_t$\\
\hline
3.0  & $>$ 0.230      \\
4.0  & 0.200 -- 0.205 \\
4.5  & 0.186 -- 0.189 \\
4.5* & 0.186 -- 0.189 \\
4.7* & 0.179 -- 0.180 \\
5.0  & 0.166 -- 0.167 \\
5.0* & 0.166 -- 0.1665 \\
5.5  & 0.1275 -- 0.130 \\
\hline
\end{tabular}
\caption{
Finite temperature transition/crossover $K_t$ for $N_F=3$
obtained on an $8^2\times10\times4$ lattice
(data with * obtained on a $12^3\times4$ lattice). 
At $\beta \leq 4.7$ clear two-state signals are observed on
both lattices.
\protect\label{tab:Kt}}
\end{center}
\end{table}


\begin{table}
\begin{center}
\begin{tabular}{ccc|ccc}
\hline
\multicolumn{3}{c|}{$m_s \approx 150$MeV} &
\multicolumn{3}{c}{$m_s \approx 400$MeV} \\
$\beta$ & $K_{ud}$ & $K_s$ &
$\beta$ & $K_{ud}$ & $K_s$ \\
\hline
3.2 & 0.2329 & 0.2043 & 3.7 & 0.2267 & 0.1692 \\
3.4 & 0.2306 & 0.2026 & 3.8 & 0.2254 & 0.1684 \\
3.5 & 0.2295 & 0.2017 & 3.9 & 0.2240 & 0.1677 \\
3.6 & 0.2281 & 0.2006 & 4.0 & 0.2226 & 0.1669 \\
4.0 & 0.2226 & 0.1964 & 4.3 & 0.2180 & 0.1643 \\
\hline
\end{tabular}
\caption{
Hopping parameters for $N_F=2+1$ simulations performed 
on $8^2\times10\times4$ and $12^3\times4$ lattices. 
$K_{ud}$ for u and d quarks is set to be equal to $K_c$
and $K_s$ for s quark is chosen so that 
$m_s \approx 150$ MeV and 400 MeV in the left and right columns,
respectively.
\protect\label{tab:1}}
\end{center}
\end{table}
\clearpage

\clearpage

\end{document}